\documentclass[aps,prb,twocolumn,showpacs, superscriptaddress,preprintnumbers,amsmath,amssymb]{revtex4}

\usepackage[dvips]{color}
\usepackage{graphicx}
\usepackage{bm}
\newcommand{ \etal }{\mbox{\sl et al. }} 
 
\newcommand{ \tbmno }{\mbox{TbMnO$_{3}$}}
 
\newcommand{ \tnmn}{\mbox{$T_N^{Mn}$}} 
\newcommand{ \tntb }{\mbox{$T_N^{Tb}$}}
\newcommand{ \Tntb }{\mbox{$T_N^{Tb}$}}
\newcommand{ \tlockmn }{\mbox{$T^{}_{s}$}} 
 
\newcommand{ \hctb }{\mbox{$H^{Tb}_{C}$}}

\newcommand{ \hca }{\mbox{$H^{a}_{C}$}} 
\newcommand{ \hcb}{\mbox{$H^{b}_{C}$}}

\newcommand{ \deltam }{\mbox{$\delta^{Mn}$}} 
 
\newcommand{ \deltatb }{\mbox{$\delta^{Tb}$}} 
\newcommand{ \tdeltatb }{\mbox{$2\delta^{Tb}$}}
 
\newcommand{ \deltamn }{\mbox{$\delta^{Mn}$}} 

\newcommand{ \degree }{\mbox{$^{\circ}$}}
 
\newcommand{ \adirection }{\mbox{a$^*$-direction}} 
\newcommand{ \bdirection}{\mbox{b$^*$-direction}} 
\newcommand{ \ppa }{\mbox{$P\|a$}}
\newcommand{ \pa }{\mbox{$P\|a$}}
\newcommand{ \ppc }{\mbox{$P\|c$}} 
\newcommand{ \pc }{\mbox{$P\|c$}} 
\newcommand{ \hpa}{\mbox{$H\|a$}} 
\newcommand{ \hpb }{\mbox{$H\|b$}} 
 
\newcommand{ \ltwo}{\mbox{$L_2$}} 
 
\newcommand{ \tn}{\mbox{$T_N$}}

\newcommand{ \bstar }{\mbox{$b^*$}}

\newcommand{ \mmsq }{\mbox{$\rm mm^2$}} 
\newcommand{\tmn}{\mbox{$\tau^{Mn}$}}
\newcommand{\ttb}{\mbox{$\tau^{Tb}$}}

\newcommand{ \TN}{\mbox{$T_N$}} 
\newcommand{ \Tn}{\mbox{$T_N$}} 
\newcommand{ \Ts }{\mbox{$T_{s}$}}
\newcommand{\bs}{\mbox{$\mathbf{b^{*}}$}}
\newcommand{\qmn}{\mbox{$\delta^{Mn}$}}
\newcommand{\qtb}{\mbox{$\delta^{Tb}$}}

\begin{document}


\title{Magnetic field induced transitions in multiferroic TbMnO$_3$ probed by resonant and non-resonant X-ray diffraction}

\author{J. Strempfer}
\email[e-mail: ]{Joerg.Strempfer@desy.de}
\affiliation{Hamburger Synchrotronstrahlungslabor (HASYLAB) at Deutsches Elektronensynchrotron (DESY), 22605 Hamburg, Germany}
\author{B. Bohnenbuck} 
\affiliation{Max-Planck-Institut f\"ur Festk\"orperforschung, 70569 Stuttgart, Germany}
\author{I. Zegkinoglou} 
\affiliation{Max-Planck-Institut f\"ur Festk\"orperforschung, 70569 Stuttgart, Germany}
\author{N. Aliouane} 
\affiliation{Hahn-Meitner-Institut, 14109 Berlin, Germany} 
\author{S. Landsgesell} 
\affiliation{Hahn-Meitner-Institut, 14109 Berlin, Germany} 
\author{M. v. Zimmermann} 
\affiliation{Hamburger Synchrotronstrahlungslabor (HASYLAB) at Deutsches Elektronensynchrotron (DESY), 22605 Hamburg, Germany}
\author{D.N. Argyriou} 
\affiliation{Hahn-Meitner-Institut, 14109 Berlin, Germany} 

\date{\today}

\begin{abstract}
Multiferroic TbMnO$_3$ is investigated using x-ray diffraction in high
magnetic fields. Measurements on first and second harmonic structural
reflections due to modulations induced by the Mn and Tb magnetic order
are presented as function of temperature and field oriented along the
$a$ and $b$-directions of the crystal.  The relation to changes in
ordering of the rare earth moments in applied field is
discussed. Observations below $T_N^{Tb}$ without and with applied magnetic
field point to a strong interaction of the rare earth order, the Mn
moments and the lattice. Also, the incommensurate to commensurate
transition of the wave vector at the critical fields is discussed with
respect to the Tb and Mn magnetic order and a phase diagram on basis
of these observations for magnetic fields $H\|a$ and $H\|b$ is
presented. The observations point to a complicated and delicate
magneto-elastic interaction as function of temperature and field.

\end{abstract}


\maketitle

\section{\label{introduction}Introduction}

Magneto-electric materials or multiferroics have stimulated much
interest both scientifically and technologically.  Scientifically
because the mechanism of coupling between ferroelectricity and magnetism is a
fundamental part in understanding the properties of
materials. Technologically because of the flexibility of controlling
states of a device with either electric or magnetic field or
both. Although modern multiferroics operate at temperatures and fields
prohibitive of direct applications they offer an exciting play ground
in order to better understand such coupled behavior.

Manganite perovskites $R$MnO$_{3}$ where $R$ is a trivalent rare earth
ion have shown an extremely rich (H,T) phase diagram of ferroelectric
and magnetic phases.  Here the frustration of magnetism in
$R$MnO$_{3}$ manganite perovskites \cite{Kim03b} offers the means to
break the chemical incompatibility between ferroelectricity and
magnetism.\cite{Hil00} The tuning of the tolerance factor with appropriate 
size $R$-ions suppresses
\Tn\ for A-type ordering of ferromagnetic Mn layers that are stacked
anti-parallel along the $c-$axis and leads to low temperature
incommensurate (IC) spin ordering for $R=$Gd, Tb and Dy.\cite{Kim03b}  
For TbMnO$_{3}$ in particular, Mn spins order below
$\TN=41$~K to form a spin density wave (SDW) with propagation vector
$\tau^{Mn}=\qmn$\bs, $\qmn\sim0.27$, while below $\Ts=29$~K the
observation of a spontaneous polarization ($\mathbf{P}$) along the
$c-$axis coincides with the onset of a transverse spiral (cycloidal) 
ordering of
Mn-spins.\cite{Ken05} At lower temperatures there is an additional
transition below $\tntb=7$~K in which Tb-spins order also incommensurately with
$\tau^{Tb}=\qtb$\bs, $\qtb=0.42$. 
Below \Ts, neutron diffraction
measurements show that the magnetic ordering is that of a transverse
spiral with Mn spins rotating within the $bc-$plane. Within the
orthorhombic P$bnm$ crystal structure of \tbmno\ ($a=5.3$\
\AA, $b=5.8$\ \AA, $c=7.4$\ \AA\ at 300 K), this ordering can be described
as:

\begin{equation}
\mathbf{M}=m_{x}\mathbf{x}+m_{y}\mathbf{y}cos(\mathbf{\tau\cdot r})+m_{z}\mathbf{z}sin(\mathbf{\tau\cdot r})
\end{equation}

where $m_{i}$ represent the magnitudes of the Mn moment along the
principal crystallographic directions $\mathbf{x,y,z}$ $(a,b,c)$ and
$\textbf{r}$ is the position of the Mn ion. For zero field $m_{x}$ is
very small or zero while below \Ts\ the orthogonal components along
$\mathbf{y}$ and $\mathbf{z}$ result in a transverse spiral ordering
of Mn-spins.\cite{Ken05} As discussed in Ref~\onlinecite{Mos06}, from
phenomenology it follows that the ferroelectric
polarization is given by:

\begin{equation}
\mathbf{P}=\gamma\chi_{e}m_{x}m_{z}[e_{x} \times \tau]
\end{equation}

where $\chi_{e}$ is the dielectric susceptibility, $\gamma$ a coupling
constant and $e_{x}$ is the spin rotation axis. Since the propagation
vector is parallel to the $b$-axis and Mn-spins rotate within the
$bc$-plane around the $a$-axis ($e_{x}\|a$), the above relation predicts the
direction of the polarization to be along the $c-$axis as indeed is
found experimentally.\cite{Kim03a} Similarly the flop of the
polarization from \pc\ to \pa\ \cite{Kim03a} that occurs under
magnetic field applied either along the $a-$ or $b-$axis, would be
expected to result from a flop of the transverse spiral from the $bc-$
to the $ab-$plane.\cite{Mos06}

The incommensurate magnetic ordering in multiferroic manganites is accompanied by
lattice deformations that result in structural super-lattice
reflections at 2$\tau$.\cite{Kim03b} The extinction condition in
reciprocal space of these second harmonic reflections are the same as for
their magnetic $\tau=\delta$\bs\ counterparts.  Phenomenologically the
nature of these reflections is magnetostrictive and arises from a
quadratic magneto-elastic coupling \cite{Wal80} between an amplitude
modulation of the magnetic moment and the lattice. However Jia
\etal\cite{Jia07} have recently discussed the
spin-lattice coupling in multiferroic manganites in terms of the
magnetostrictive, orbital and Dzyaloshinskii-Moriya (DM) interactions
(or $S_{i}\times S_{j}$ spin-current terms \cite{Kat05}) on the
basis of the electronic configurations of insulating manganites. From
these results it follows that for a spin density wave (SDW), 2$\tau$ lattice
reflections are expected to be observed, while for a circular transverse spiral
($m_{y}=m_{z}$), 2$\tau$ reflections are
suppressed.\cite{Jia07}  If the transverse spiral
becomes conical ($m_{x}\neq0$) it is possible to observe structural
reflections also at $\tau$.\cite{Jia07} This means that
first harmonic reflections can have a mixed magnetic and lattice
character.  Indeed we have observed in magnetic fields the \qmn\bs\
reflection in non-resonant x-ray diffraction
experiments \cite{Ali06} for ${\mu_0H}>1$~T. Therefore the
investigation of incommensurate reflections in multiferroic manganites using 
X-ray diffraction
can provide valuable information on the type of magneto-elastic coupling
that can be active in these multiferroics.

More recently it was shown that in TbMnO$_{3}$ the magnetic ordering
of Tb- and Mn-spins are highly coupled below \Ts.\cite{Pro07b} While
for $T>\Ts$ the magnetic wave vectors for Tb and Mn are locked so that
$\tau^{Tb}=\tau^{Mn}$,\cite{Ken05} below \Tntb\ it is found that
$\tau^{Tb}$ and $\tau^{Mn}$ lock-in 
to wave vectors whose magnitudes are rational fractions 3/7\bs\ and
2/7\bs\ respectively, 
while the wave vectors hold the relationship $3\tau^{Tb} -
\tau^{Mn} = 1$.  This novel matching of wave vectors can be described
within the frustrated Anisotropic-Next-Nearest-Neighbor-Ising (ANNNI) 
model coupled to a periodic external field
produced by the Mn-spin order, as detailed in Ref.~\onlinecite{Pro07b}. 
Within this model the
$\tau^{Tb}=\tau^{Mn}$ behavior is recovered while the
$\tau^{Tb}=3/7\bs$ and $\tau^{Mn} = 2/7\bs$ regime is stabilized by
an optimal ordering of 6 domain walls in the Tb spin-density wave,
superimposed on the Mn-order. This model further shows that the
ordering of $\tau^{Tb}=3/7\bs$ is energetically more favorable than
the simple $\uparrow \uparrow \downarrow \downarrow$ with
$\tau=1/2\bs$ found at low temperatures for
DyMnO$_{3}$.\cite{Fey06}

In this paper we report on x-ray
diffraction measurements of the lattice deformation in TbMnO$_{3}$ as
a function of temperature and magnetic field applied parallel to the $a-$ and
$b-$axis. The paper is organized in the following way. In section II we
present the experimental setups used for the present investigations. 
In section III we report on zero field measurements as a function of 
temperature for selected 2$\tau$ reflections. In section IV, 
we present measurements with magnetic field applied along the \adirection\
and in section V, with magnetic field applied along the \bdirection.
The phase diagram resulting from our data is discussed in section VI.

In this paper we follow the convention of labeling incommensurate
magnetic reflections from Bertaut's representational
theory.\cite{bertaut} Here the magnetic ordering of Mn-spins within
space group P$bnm$ and \tmn=0.27\bstar\ can be described by four
irreducible representations ($\Gamma$) that consist of four modes
(labeled as A, C, F and G).\cite{bertaut} The modes correspond to
magnetic superlattice reflections that occur in different Brillouin
zones with extinction conditions as follows; $A$ for $h+k$=even,
$l$=odd; $G$ for $h+k$=odd with $l$=odd; $F$ for $h+k$=even and
$l$=even; and $C$ for $h+k$=odd and $l$=even, where $h,k,l$ are Miller
indices. Each mode here describes the Mn-spin polarization along one
unique principle crystallographic axis.\cite{bertaut} It is shown that
the Mn transverse spiral ordering in \tbmno\ is described by two
irreducible representations $\Gamma_{2}\times\Gamma_{3}$ where,
$\Gamma_{2}=(C_{x},F_{y},A_{z})$ and
$\Gamma_{3}=(G_{x},A_{y},F_{z})$.\cite{Ken05} The ordering of the
Mn-spins is described by two A-modes, $A_{y},A_{z}$.\cite{Ken05} Using
a combination of unpolarized and polarized neutron diffraction and
resonant magnetic X-ray scattering it has been shown that the
remaining modes arise from the ordering of Tb-spins.\cite{Ken05,Ali08}
In the following, the notation $\tau^{Mn/Tb}=(0, \delta^{Mn/Tb}, 0)$ for 
first harmonic and $2\tau^{Mn/Tb}=(0, 2\delta^{Mn/Tb}, 0)$ for 
second harmonic Mn and Tb incommensurate 
superlattice reflections is used, with
wave vector $\tau$ and wave number $\delta$. In the intermediate and
the commensurate phases, numerical values for wave numbers are used.

\section{\label{experimental}Experimental}

\begin{figure}[htb!]
\begin{center}
\includegraphics[width=8cm]{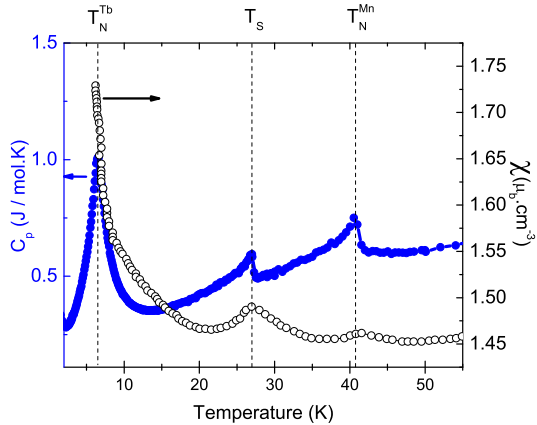}
\caption{(Color online) 
Temperature dependence of the specific heat (C$_{p}$) and
magnetic susceptibility $\chi$ of a small single crystal from the
TbMnO$_{3}$ boule. For the magnetization
measurements a field of 0.01~T was applied along the $c-$axis.  The
C$_{p}$(T) and $\chi$(T) data indicate three successive phase
transitions at 7~K, 27~K and 41~K corresponding to
\Tntb, \Ts, \TN, respectively. } \label{chara}
\end{center}
\end{figure}

The \tbmno\ crystals used in our experiments were grown at the
Hahn-Meitner-Institute in Berlin using the floating zone technique
under Ar atmosphere. The crystals cut from the crystalline boule
show an excellent crystal quality with a mosaic spread of
0.016\degree\ of the (0\ 2\ 0) reflection (FWHM). 
In Fig.~\ref{chara} we show measurements of the
temperature dependence of heat capacity and magnetic
susceptibility from a small single crystal cut from the same
crystalline boule. The data shows three successive transitions with
decreasing temperature at $\TN=41$~K, $\Ts=28$~K and $\Tntb=7$~K. 
The measurements and transition temperatures are in
good agreement with published measurements,\cite{Kim03a} indicating a
high quality crystal.

The experiments were performed both at the beamline BW5 at the
Hamburger Synchrotronstrahlungslabor (HASYLAB) and at the beamline X21
at the National Synchrotron Light Source (NSLS) at Brookhaven National
Laboratory.

At HASYLAB the experiment was conducted at a photon energy of 100keV
in horizontal scattering geometry. The sample was mounted in a
Cryogenics superconducting cryomagnet with horizontal field up to
10~T. The beam was monochromized by a (111)-SiGe gradient crystal. A
second SiGe gradient crystal was used as analyzer to suppress background.
The single crystalline \tbmno\ sample had a size of 2x3x0.6~mm$^3$ and
was measured in transmission geometry, which means the true bulk of
the crystal is probed. The sample thickness in the direction of the
beam was of the order of the absorption length
at this high x-ray energy and was thus optimum for obtaining maximum
scattering intensity. The crystal was aligned with the {\em bc}-plane
in the horizontal diffraction plane in order to access $(0,\ k,\ l)$
reflections. Using this setup, measurements were performed with field
\hpb.

At NSLS, the experiment was conducted in the hard x-ray regime at
9.5~keV as well as at the Tb \ltwo\ absorption edge with a 
photon energy of 8.252~keV. The sample was mounted in
a 13~T Oxford cryomagnet with vertical magnetic field. 
With this setup, measurements with field \hpa\ were conducted, with
the {\em bc}-plane oriented in the horizontal diffraction plane. 

A $(0\ 0\ 2)$ graphite
analyzer was used for background reduction for the non-resonant
measurements.  As detector, a field insensitive Avalanche Photodiode
(APD) was used.

For measurements performed at the absorption edge polarization
analysis was performed using the $(0\ 0\ 6)$ reflection of the
graphite analyzer. The scattering geometry was $\pi-\sigma'$ with the
analyzer at an angle of $2\theta_{(006)}=84.6^{\circ}$ implying a
leakage of less then 1$\%$ from the $\pi-\pi'$ channel.

The sample used here was a crystal with polished c-surface and a size of
about 3x3~\mmsq\ in order to obtain high diffracted intensities in
Bragg geometry.

All results presented in this study measured away from the resonances
at the absorption edges are due to pure charge scattering.  This can
safely be assumed since the non-resonant magnetic scattering
cross-section is about six orders of magnitude weaker than the charge
scattering cross-section. Besides this, the magnetic signal of the
observed reflections is further reduced due to the small magnetic form
factor at the high Q-values we investigate here.

\section{\label{zerofield} Measurements at zero field}

\begin{figure}
 \includegraphics[width=8cm]{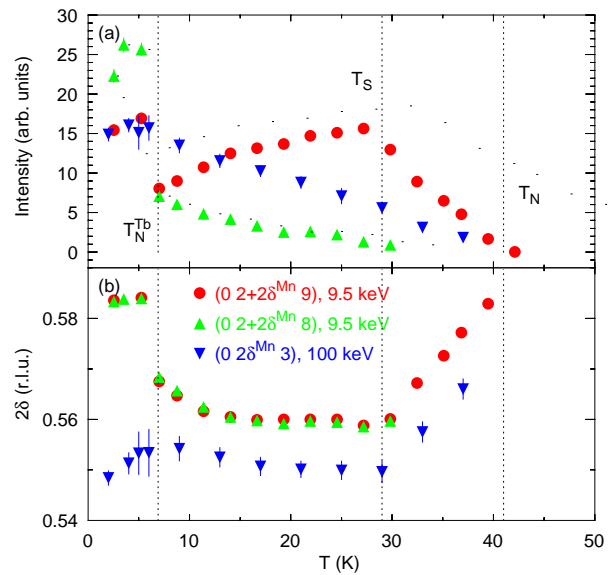}
 \caption{\label{fig:tdep_h0} (Color online) 
Temperature dependence at zero field and 9.5~keV photon energy of
 the (a) intensity and (b) k-component of the wave vector of the
 A-mode (0, 2+2\deltam, 9) and F-mode (0, 2+2\deltam, 8)
 reflections. The respective temperatures \tn, \tlockmn\ and \tntb\ 
are indicated by dashed lines. In addition data for the A-mode
(0, 2\deltam, 3) reflection measured at 100~keV are plotted.
Due to a systematic offset the absolute wave numbers are not accurate.}
\end{figure}

In Fig.~\ref{fig:tdep_h0} we show the dependence of the intensity and
wave vector with temperature of the A-mode (0,\ 2+2\deltamn,\ 9) and
the F-mode (0,\ 2+2\deltamn,\ 8) reflections measured using a photon
energy of 9.5~keV.  From this data it is clear that the variation of
intensity of the two reflections 
with decreasing temperature is substantially different
and reflects the origin of these reflections. The A-mode reflects the
lattice modulation that arises purely from the ordering of Mn-spins,
while the F-mode reflects the lattice distortion that arises from the
induced ordering of Tb-spins with $\tau^{Tb}=\tau^{Mn}$. For the
A-mode reflection with decreasing temperature below \TN\ we find a linear increase in
its intensity up to \Ts\, while below \Ts\ its intensity decreases
with further cooling down to \tntb.  
As Tb-spins order, the intensity
of these reflections rapidly increases for $T<\tntb$.  This is
in sharp contrast to the F-mode reflection where the intensity shows a
smooth increase with decreasing temperature down 
to \tntb\ while a similar jump in
intensity as for the A-mode reflections is observed below \tntb.  In
Fig. ~\ref{fig:tdep_h0} we show also data measured from the A-mode
(0,\ 2\deltamn,\ 3) reflection 
using 100~keV x-rays.  The behavior here is different to the 9.5~keV data. 
The intensity
increases smoothly with temperature down to \tntb, below which it
saturates. No jump in intensity nor a jump of the wave vector at \tntb\ is observed here. A possible scenario for the behavior of this reflection
is presented at the end of this section.

\begin{figure}
 \includegraphics[width=8cm]{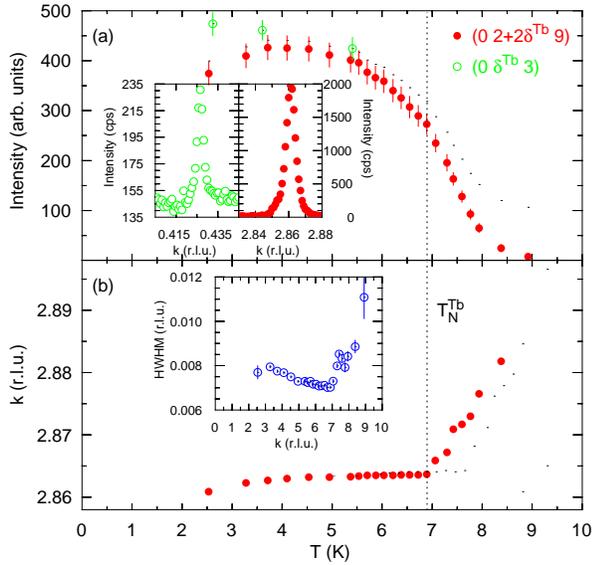}
 \caption{\label{tdep_tb_0t} (Color online)
Tb superlattice reflection as function of 
temperature. (a) shows the intensity of the $(0,\ 2+\tdeltatb,\ 9)$ and
the $(0,\ \deltatb,\ 3)$. (b) shows the variation of the propagation
vector of the $(0,\ 2+\tdeltatb,\ 9)$ superlattice reflection. In the inset,
a scan over the first and second harmonic 
peak along the \bdirection\ is shown.}
\end{figure}

\begin{figure}
 \includegraphics[width=8cm]{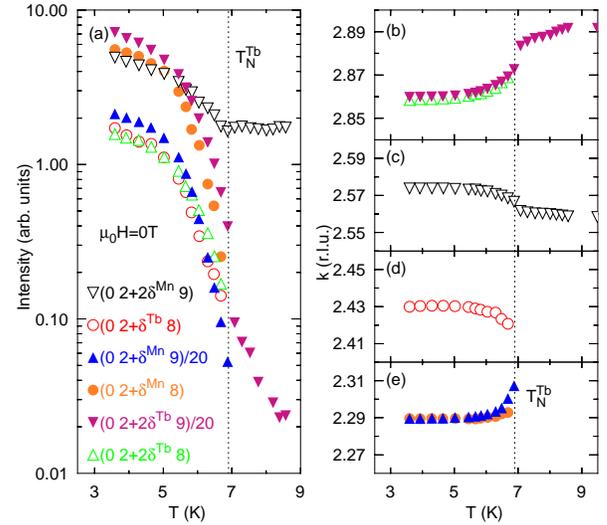}
 \caption{\label{tdep_0t_tb2} (Color online) 
Temperature dependence of (a) intensities and (b-e) wave number of the first and  second harmonic structural Mn- and Tb-superlattice 
reflections in the Tb ordered phase $T<\tntb$ at zero field.  }
\end{figure}

In Fig.~\ref{tdep_tb_0t} we show detailed measurements of the (0,
$2+2\qtb$, 9) and (0, \qtb, 3) reflections below $\tntb=9$~K that
describe the lattice modulation associated with the change of the
Tb-ordering from $\tau^{Tb}=\tau^{Mn}$ to
3$\tau^{Tb}-\tau^{Mn}=$1.\cite{Pro07b} The intensity of the 2\ttb\
reflection shows a typical order parameters behavior with cooling
below \Tntb\ that follows the ordering of Tb-spins. Below 6~K we
can also measure the first harmonic (0, \qtb, 3) reflection
(Fig.~\ref{tdep_tb_0t}(b)) that is visible with non-resonant x-ray
scattering showing that below \Tntb\ there is a structural component
to the first harmonic magnetic reflection. This is supported by the observation
of a strong first harmonic Tb reflection shown in 
Fig.~\ref{tdep_0t_tb2}.

The variation with temperature for the wave vectors 2$\tau^{Mn}$ and
2$\tau^{Tb}$ is shown in Fig.~\ref{fig:tdep_h0}(b) and
~\ref{tdep_tb_0t}(b) respectively.  We find that the value of 2\qmn\
rapidly decreases on cooling through the SDW regime as noted earlier
and shows a weaker temperature dependence below \Ts.\cite{Ken05}
However at \tntb\ we note a substantial increase in the value of
2\qmn, that tracks closely the rapid changes in \qtb\ between 9 and
7~K, while below 7~K the values of both incommensurabilities remain
relatively temperature invariant. Below \tntb\ we confirm the
observation that the wave vectors for Mn- and Tb-spin order are
coupled. Indeed 
the wave vectors of the (0, 2+2\qtb, 9) and (0, 2+2\qtb, 8)
as well as the (0, 2+\qmn, 8) and (0, 2+\qmn, 9) 
in Fig.~\ref{tdep_0t_tb2}b-e are following the
relation $3\ttb-\tmn=1$.
Below 5~K the respective wave numbers
approach the values of $\qmn=2/7$ and $\qtb=3/7$
within 0.002 accuracy, as was shown to be the case in
Ref.~\onlinecite{Pro07b}.

We note here that we found surprising differences in the change of the
value of \qmn\ through the transition at \tntb.  Through this
transition Kenzelmann \etal (Ref.~\onlinecite{Ken05}) report a jump in \qmn\ of
$\sim10^{-3}$ using single crystal neutron diffraction.  This value is
similar to the neutron measurements reported in Ref.~\onlinecite{Ali08} 
as well as X-ray diffraction measurements using
100~keV x-rays shown for the (0, 2\qmn, 3) reflection in
Fig.~\ref{fig:tdep_h0}(b).  This is in contrast to the $\sim10^{-2}$
change we find across \tntb\ using 9.5~keV x-rays in this
work (Fig.~\ref{fig:tdep_h0}(b)) as well as in
Ref.~\onlinecite{Pro07b}. In the case of the 9.5~keV experiments we
probe few $\mu m$ of the surface of the crystal while in the neutron
and high energy x-ray experiments we probe the bulk of the sample.
This one order of magnitude difference in the change of \qmn\ across
\tntb\ suggests that strain effects at the crystal surface allows the
lock-in of the wave vectors to values of rational fractions within
$\Delta\delta\sim10^{-2}$, while the unstrained bulk appears to
modulate the values of the wave vectors below \tntb.

We now turn our attention to the intensity variation with temperature
of the A- and F-mode 2$\tau^{Mn}$ reflections shown in
Fig.~\ref{fig:tdep_h0}. The behavior of the A-mode (0,\ 2+2\deltamn,\
9) reflection can be understood in terms of the changes in the
magnetic ordering of Mn-spins. In the SDW regime between $\Ts <T< \TN$
the increase in intensity of the 2$\tau^{Mn}$ A-mode arises from the
quadratic magneto-elastic coupling in which the amplitude of the
lattice displacements varies linearly with the increase in the size of
ordered Mn moment.\cite{Wal80,Jia07} The decrease in
intensity below \Ts\ is ascribed to the development of a perpendicular
component that alters the collinear SDW to a transverse
spiral.\cite{Jia07} For a perfect spiral with $m_{y}=m_{z}$, 
the 2\tmn\ reflection would be completely
suppressed,\cite{Jia07} however our data shows that at a
temperature just above \Tntb\ the 2\tmn\ reflection is still observed
indicating that the spiral remains elliptical ($m_{y}\neq
m_{z}$). Indeed this is confirmed by neutron diffraction by measuring
directly the values of $m_{y}$ and $m_{z}$ to be 3.9 and 2.8 $\mu_{B}/Mn$ 
respectively.\cite{Ken05} At \Tntb\ the
intensity of the 2$\tau^{Mn}$ A-mode jumps dramatically. At first
glance this rise in intensity may suggest changes in the
magneto-elastic coupling of Mn- and Tb- sublattices. However, more
likely the jump in intensity below \tntb\ arises from the matching of
the Mn and Tb wave vectors as described above. Since we know that below \Tntb\
we have significant structural contribution to the magnetic 
$\tau^{Tb}$ reflections it is most likely that here the $\tau^{Tb}$ intensity
is superimposed on a
2$\tau^{Mn}$ reflection below \tntb\ and thus providing dramatic changes
in intensity. 

More puzzling is the difference in behavior
of the A-mode reflections measured with 9.5~keV and 100~keV
x-rays also shown in Fig.~\ref{fig:tdep_h0}, where the 100~keV data show a linear variation in
intensity with cooling from \Tn\ down to \tntb\ in contrast 
to the behavior observed for 9.5~keV photons. Here again, we believe this has to do with surface versus bulk properties of the material, since 
also measurements performed with bulk sensitive neutron scattering show
the behavior of the A-mode reflections we observe here with high-energy
x-rays.\cite{Kaj04,Ken05}

\section{\label{fieldpa}Field orientation $H\|a$}

\begin{figure}
 \includegraphics[width=8cm]{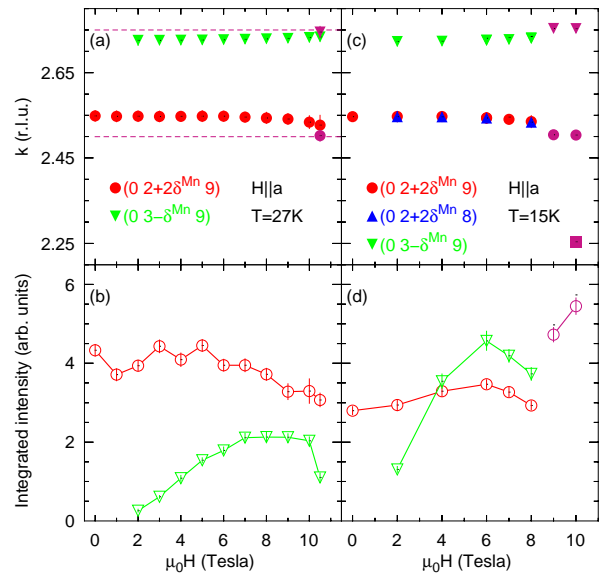}
 \caption{\label{hdep_27k_a} (Color online) Magnetic field dependence of (a) wave vector and 
(b) intensity of the second harmonic 
$(0,\ 2+2\deltamn,\ 9)$ (circles)
and first harmonic $(0,\ 3-\deltamn,\ 9)$ (triangles)
superlattice reflection at a sample temperature of $T=27$~K for \hpa. 
Magnetic field dependence of (c) wave vector and (d) intensities of the CM and
IC superlattice reflection for magnetic field \hpa\ at
a sample temperature $T=15$~K. At $\mu_0\hca=9$~T the
reflections lock in at CM positions. The horizontal dashed
lines show the half and quarter 
integer positions of the CM first and second harmonic 
reflections above \hca.}
\end{figure}

The application of magnetic field along the $a-$axis results in
strong modulations of the magnetic order of Tb- and Mn-spins with the
consequence of affecting significantly the ferroelectric
properties. The most notable change with field is the flop of the
direction of the spontaneous polarization from \ppc\ to \ppa.
Phenomenologically this is attributed to the flop of the Mn spin
spiral from the $bc-$plane to the $ab-$plane.  This flop is
accompanied by a change of \tmn\ to a commensurate value of 1/4 at a
critical field \hca\ of approximately 10~T at
2~K.\cite{Ali06} In this section we focus predominantly 
on field induced
magneto-elastic transitions that occur below \hca\ and give us an
insight into both the magnetic ordering and magneto-elastic coupling
in TbMnO$_{3}$.

\subsection{\label{mnordered_hpa} Phase region $\tntb<T<\tnmn$}

We measured the field dependence for \hpa\ for $\tau$ and 2$\tau$
reflections at $T=27$, 15 and 9~K. In Fig.~\ref{hdep_27k_a}a-b and
~\ref{hdep_27k_a}c-d, we show the field dependence at $T=$~27 and 15~K,
below \Ts, for the A-mode 2\tmn\ reflection $(0,\ 2+2\delta^{Mn},\ 9)$
and the F-mode $(0,\ 2+2\delta^{Mn},\ 8)$ reflection for 15~K only. In
this regime Tb- and Mn- spins are ordered with the same wave vector
(\tmn=\ttb). The \tmn\ and 2\tmn\ reflections for the C- and G-modes
are too weak to be observed at these two temperatures.

The intensity of the 2\tmn\ A- and F-modes for $T=$27 and 15~K is
essentially constant with increasing field up to \hca\ suggesting that
field does not change either the ellipticity of the Mn spin-spiral or
the magneto-elastic coupling associated with it. Surprisingly above 2~T
for both 27 and 15~K we observe the \tmn\ G-mode $(0,\ 3-\delta^{Mn},\
9)$ reflection.  The wave number of this reflection is \qmn=0.275 at
$\mu_0H=2$~T, in agreement with the incommensurability 2\qmn=0.55 of
the 2\tmn\ reflection. The intensity of this reflection shows a linear
increase with field above 2~T and saturates at approximately 6~T
(Fig.~\ref{hdep_27k_a}b and d).
Similar measurements as a function of field at 9~K just above \Tntb\ show the
very same behavior as described above for the same reflections.
Additional scans (not shown) demonstrate that only \tmn\ G-
and C-modes are observed above 2~T, while the 2\tmn\ A- and F-modes are always
measurable irrespective of field. 
The observation of \tmn\
reflections in fields larger than 2~T may have a number of possible
reasons. The emergence of a first harmonic reflection may be
thought at first glance as trivial as the coupling of the lattice to
$H$ becomes linear.  However Jia \etal\ suggest that a linear
behavior to the spin-lattice coupling may occur in the case when the
transverse spiral becomes conical i.e. $m_{x}\neq
0$.\cite{Jia07} If this was indeed true then we would
expect to observe the \tmn\ reflection of an A-mode as such reflection
describe the Mn spiral ordering.  However according to our observations
the first harmonic reflections in field are found
for G- and C-modes suggesting possible changes to the Tb-spin ordering
with magnetic field.  Indeed magnetization measurements in
Ref.~\onlinecite{Kim05} suggest the ferromagnetic alignment of
Tb-spins with \hpa\ for T=15 and 9~K.  Therefore the emergence of \tmn\
reflections with field most likely is associated with changes in the
Tb-ordering as opposed to changes in the magneto-elastic coupling.

\subsection{\label{tbordered_hpa} Phase region $T<\tntb$}

\begin{figure}
 \includegraphics[width=8cm]{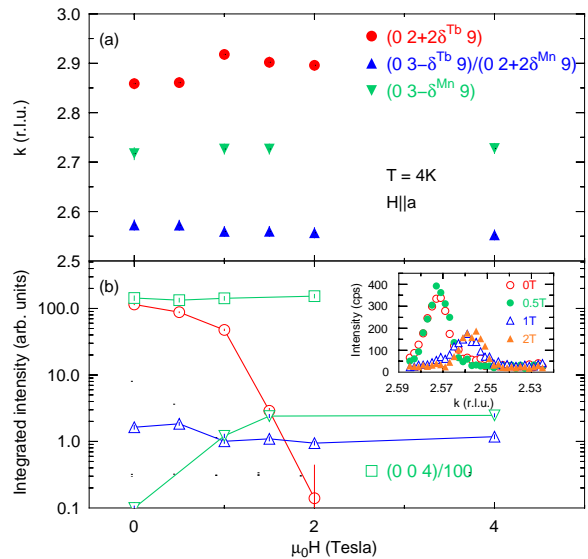}
 \caption{\label{hdep_4k_a} (Color online) 
Field dependence at $T=4$~K of the terbium
and manganese superlattice reflections with magnetic field \hpa. 
(a) peak positions (closed symbols) and (b) the corresponding 
intensities (open symbols) are shown. Open rectangles
show the (0, 0, 4) main Bragg intensity divided by a factor 100. 
The inset in (b) shows the peak profiles of the (0, 3-\deltatb, 9) 
and (0, 2+2\deltatb, 9) reflections for different fields.
}
\end{figure}

We now turn our attention to the behavior of the incommensurate reflections as a
function of field below \Tntb\ and remind the reader that in this
regime at zero field the Tb and Mn magnetic ordering is coupled so
that $3\ttb-\tmn=1$. As shown Fig.~\ref{hdep_4k_a}  we find  \ttb\ and the A-mode \tmn\  for $\mu_0H=1$~T and $T<\Tntb$.  Increasing magnetic field results in the strong attenuation of
the 2\ttb\ (0, 2+2\qtb, 9) reflection and its complete suppression
above $\mu_0H>2$~T, consistent with the FM ordering of Tb-spins as
indicated by neutron diffraction.\cite{Ali06} 

For
intermediate field values ($0<H<2$~T) the 2\ttb\ reflection vanishes from the $
2\qtb=0.86$ position and seems to shift to ~0.90 
(Fig.~\ref{hdep_4k_a}). On the other hand
resonant scattering from the Tb \ltwo\ absorption edge indicates a shift
of the first harmonic \ttb\ reflection to 0.364 indicating a shift of the
2\ttb\ reflection to the 0.727 position, where also a peak appears at
$\approx$1~T as will be shown later. This reflection nevertheless stays also 
above a field of 2~T, which indicates that we deal here with the \tmn\ 
reflection observed also at temperatures $T>\tntb$. 
The shift observed at 1~T is coupled to a discontinuous change in the
2\qmn\ reflection from 0.572 to 0.56 also for the same field values as
shown in the inset of Fig.~\ref{hdep_4k_a}(b). This again is due to the 
superposition of the \ttb\ intensity on the 2\tmn\ reflection at low fields
and a pure 2\tmn\ reflection at higher fields.

\begin{figure}
 \includegraphics[width=8cm]{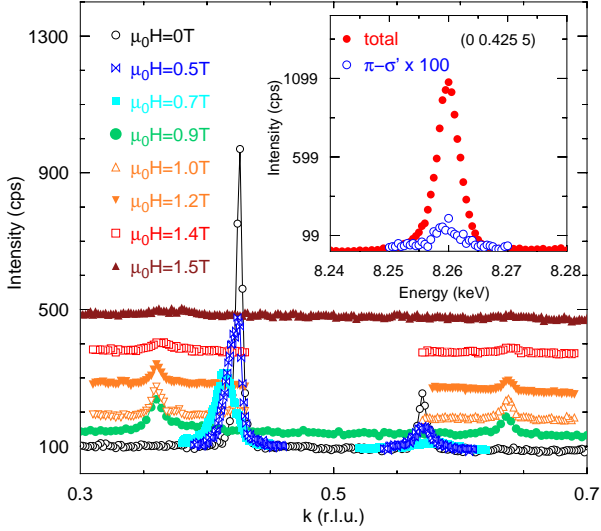}
 \caption{\label{hdep_4k_resonant} (Color online)
K-scans of the (0, \deltatb, 5) and 
(0, 1-\deltatb, 5) reflection as function of field \hpa\ at a sample 
temperature $T=4$~K without polarization analysis. The
inset shows an energy scan over the Tb \ltwo\ absorption edge with and 
without polarization analysis in the $\pi-\sigma'$ channel. The total signal only shows the resonant 
intensity. The fluorescence background is subtracted using an energy scan
performed off the reflection position.
}
\end{figure}

To investigate the transition to this intermediate phase we performed
resonant X-ray diffraction measurements with and without polarization analysis 
in the $\pi-\sigma '$ polarization channel as
a function of \hpa, by tuning the photon energy to the Tb \ltwo-edge at
8.252~keV. At this photon
energy, the background was significantly higher than for the
experiments at 9.5~keV and prohibited the observation of any
non-resonant scattering signal. However, a strong resonant magnetic signal was
observed at $T=4$~K for the first harmonic Tb reflections $(0,\
k\pm$\qtb$,\ l)$ with $l={\rm odd}$ ( Tb magnetic A- and C-modes).
The measurements were conducted with analyzer used to reduce
background (PG(002)) and as polarization analyzer (PG(006)) for the
$\pi-\sigma'$ channel. In the inset of Fig.~\ref{hdep_4k_resonant},
energy scans with and without polarization analysis are shown. The
small intensity in the $\pi-\sigma'$-channel is due to leakage since
the analyzer angle is not exactly at $2\theta=90^{\circ}$ but at
$2\theta_{(006)}=84.6^{\circ}$. All intensity is thus scattered in the 
$\pi-\pi'$ channel. According to the resonant magnetic scattering 
cross section, 
this is the case if the magnetic Tb moment is aligned along the
\adirection\ perpendicular to the scattering plane, 
consistent with results in Ref.~\onlinecite{Man07}.

In Fig.~\ref{hdep_4k_resonant} scans over the (0, \qtb, 5) and (0,
1-\qtb, 5) positions with $\qtb=0.43$ are shown as function of field. As
the field is increased, the intensity of these reflections 
decreases and the wave number \qtb\ slightly shifts as observed in the
non-resonant experiment.  At $\mu_0H=0.9$~T, this resonant reflection
shifts discontinuously from $\qtb=0.428(2)$ ($\mu_0H=0$~T) to
$\qtb=0.363(1)$ ($\mu_0H=1$~T). While the wave number of this reflection
is invariant in field, its intensity is rapidly reduced with
increasing field and vanished above $\hpa=2$~T, when Tb-spins show a FM
ordering ordering.\cite{Ali08} Our resonant and non-resonant
measurements demonstrate that this intermediate phase is stable
between $\hpa=0.9$ and 2~T.

The discontinuous transition we find here at 1~T suggests a change in
the coupling of the Mn and Tb ordering. In this new regime the values
of \qtb\ and \qmn\ are close to the rational fractions 4/11 and
3/11, respectively, while the wave vectors hold the relationship of
$2\ttb+\tmn=1$ with 0.001 accuracy. Modeling of the Mn- and Tb-spins
using an ANNNI model it was shown that such a state is stabilized as
the Tb-SDW acquires a homogeneous component via $\chi H$ and couples
to the quartic term of the Landau expansion.\cite{Pro07b} Interestingly
the change in the coupling between Mn and Tb magnetic ordering
characterizes a region in which the value of \pc\ increases by
$\sim$30\% up to $\hpa=2$~T.

For $\hpa>2$~T the FM alignment of
Tb-spins results in the melting of
the incommensurate Tb-ordering.\cite{Ali08,Kim05,Ali06} 
This melting is associated with a
decrease of \pc\ to values 30\% smaller than the maximum polarization
at zero field. However \tmn\ and 2\tmn\ reflections are observed above
2~T and discontinuously shift to values of 1/4 and 1/2 respectively at
\hca\ (not shown).

\section{\label{fieldpb}Field orientation $H\|b$}

\begin{figure}
 \includegraphics[width=8cm]{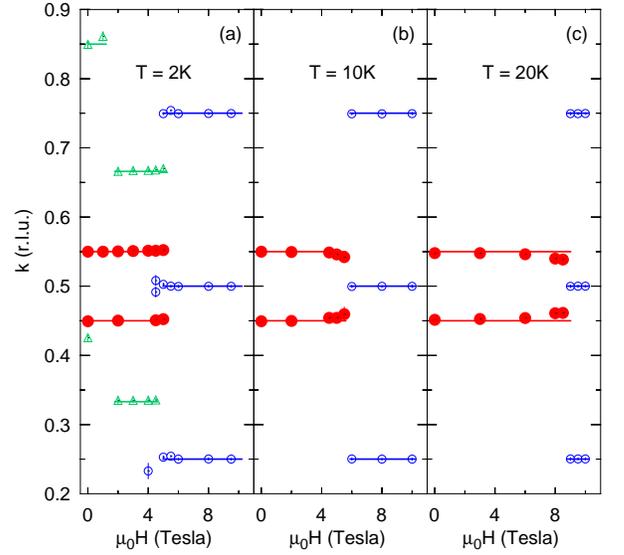}
 \caption{\label{fig:hdep_2_10_20} (Color online) Incommensurability 
$\delta$ of the 
superlattice reflections 
(0, 2\deltamn, 4) and (0, 1-2\deltamn, 4) below \hcb\ (closed circles)
and (0, 0.25, 4),(0, 0.5, 4) and (0, 0.75, 4) above \hcb\ 
(open circles), measured at 
(a) $T=2$~K, (b) $T=10$~K and (c) $T=20$~K as function of magnetic field \hpb. 
At $T=2$~K in addition (0, \deltatb, 5) and (0, 2\deltatb, 5) reflections
are shown below and above \hctb\ (open triangles).
}
\end{figure}

Whereas the \hpa\ measurements were performed with photon energies of 9.5~keV
that allows to probe only the first few $\mu m$ of the crystal,
the measurements with \hpb\ are performed with high photon energies
of 100~keV with which the true bulk of the crystal is probed.
For the \hpb\ geometry the sequence of magneto-structural transitions
towards the polarization flop transition is significantly different
than what we have found for the \hpa\ configuration.  Although we find
that the polarization flop again corresponds to a first order transition
to a commensurate phase with \qmn=1/4, the sequence of transitions
associated with Tb-ordering differs from the \hpa\ case.
A summary of our results with respect to the wave numbers \qmn\ and
\qtb\ is shown in Fig.~\ref{fig:hdep_2_10_20}a-c for $T=2$, 10 and 20~K
respectively. The transition to $\qmn=1/4$ is clearly evident from the data
while \hcb\ decreases with decreasing temperature to values that agree well 
with the observation of the flop in polarization from \ppc\ to
\ppa.\cite{Kim03a,Kim05,Ari05}

\subsection{\label{mnordered_hpb} Phase region $\tntb<T<\tnmn$}

\begin{figure}
 \includegraphics[width=8cm]{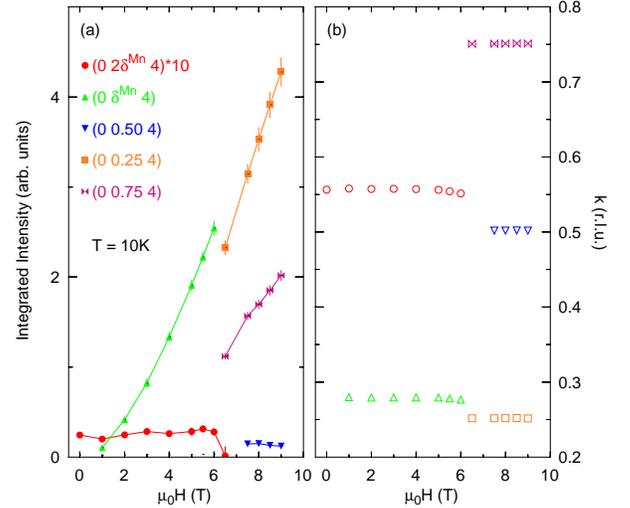}
 \caption{\label{fig:hdep_10k_new} (Color online) 
(a) Intensities as function of field of the IC $(0,\ \deltamn,\ 3)$ and 
$(0,\ 2\deltamn,\ 4)$ reflections below \hcb\ and the CM
$(0,\ 0.25,\ 4)$, $(0,\ 0.5,\ 4)$ and
$(0,\ 0.75,\ 4)$ above \hcb. (b) The respective wave vectors (0 k 4) for the 
intensities in (a) are shown. }
\end{figure}

In the region of $\Tntb<T<\TN$, we find only reflections with 2\tmn\ as
expected from previous measurements and indicative of quadratic
magneto-elastic coupling.  However above $\hcb>1$~T we find also first
harmonic reflection such as the F-mode (0, \qmn, 4).  This reflection
that arises from the ordering of Tb-spins with $\ttb=\tmn$ shows a
linear increase in intensity with fields up to \hcb\ indicating changes in the magnetic ordering of Tb-spins. In contrast its
second harmonic counterpart (0, 2\qmn, 4) remains constant in intensity
up to \hcb\ (Fig.~\ref{fig:hdep_10k_new}a).  This linear increase over
the whole field range is different from the behavior found for \hpa\
(section~\ref{fieldpa}), where 
the intensity of these reflections increases initially
but then decreases again towards \hca. 

As we approach the critical field \hcb\ the value of 2\tmn\ begins to move
towards 1/2, while at \hcb\ we find a first order phase
transition where the intensity of the incommensurate reflection decreases with field 
and vanishes as the commensurate reflection appears
(Fig.~\ref{fig:hdep_10k_new}b). With 
decreasing temperature the value of \hcb\ also decreases as shown in
Fig.~\ref{fig:hdep_2_10_20}a-c. These changes to \tmn\ and 2\tmn\
F-mode reflections have to be attributed to changes in the magnetic
structure associated with Tb-spins. We
speculate that at this high temperature the application of magnetic
field may result in the polarization of the Tb-spins along the
$b-$axis. For this field direction there is no evidence of ferromagnetic ordering of Tb-spins as found for \hpa.\cite{Kim03a, Kim05,Ali08}

\begin{figure}
 \includegraphics[width=8cm]{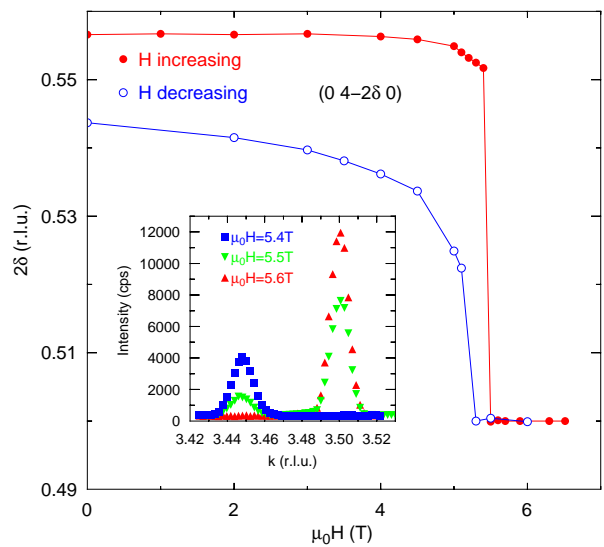}
 \caption{\label{hysterese} (Color online) Hysteresis of the 
superlattice reflections measured at $T=10$~K. Solid circles show the data
measured with increasing field, open circles the data measured with
decreasing field. The inset shows the superlattice reflections at the phase
transition for increasing field.}
\end{figure}

Detailed measurements of the transition to the commensurate phase were
made at 10~K by tracking the lattice modulation associated with the
F-mode structural peak (0, $4-2\qmn$, 0). These measurements, shown in
Fig.~\ref{hysterese}, were taken by cooling the sample to 10~K in zero
field and applying field up to $\hpa=6.5$~T and then down to 0~T while the
data were measured. These data show a hysteresis in terms of
field and wave number 2\qmn\ that tracks the hysteresis in the
polarization flop from \ppc\ to \ppa.\cite{Kim03a, Kim05}  In terms of
magnetic field the hysteresis is as small as $\sim$0.2~T.  However in terms
of wave number the effect is more pronounced. For increasing field the
wave number remains at a value of $2\qmn=0.557$ almost up to \hcb,
before it locks to the commensurate 
value of 1/2. For decreasing field, a rapid 
increase in 2\qmn\ just below \hcb\ and a significantly
lower value of $2\qmn=0.543$ at $\mu_0H=0$~T is observed. Interestingly, these
lower values suggest $\qmn= 0.272$ which is close to the rational
fraction 3/11 as found for the case of \hpa\ and 9.5~keV photons below
\tntb. In the inset of Fig.~\ref{hysterese}, scans over the
superlattice reflections for increasing field around \hcb\ are
displayed. It shows the coexistence of the incommensurate and
commensurate phase right at the transition at $\mu_0\hcb=5.5$~T. The
intensities of these reflection in the incommensurate phase for
increasing and decreasing field show exactly the same values
while only their positions show a hysteresis.

\subsection{\label{tbordered_hpb} Tb-ordered phase region $T<\tntb$}

\begin{figure}
 \includegraphics[width=8cm]{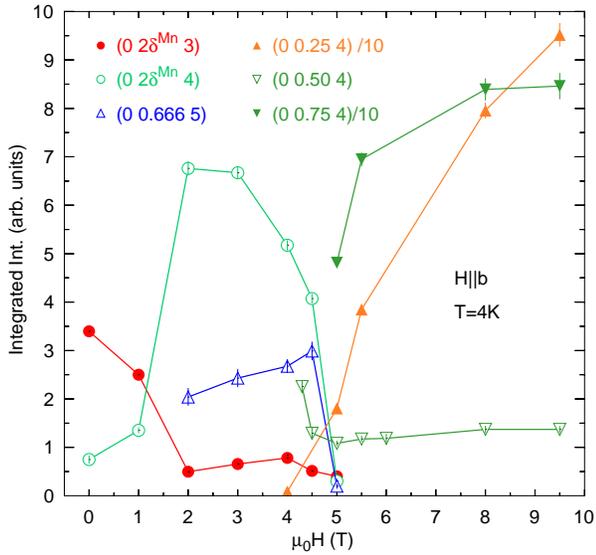}
 \caption{\label{fig:hdep_2k_int} (Color online) 
Intensities of the $(0\ 0.55\ 3)$, 
$(0\ 0.55\ 4)$, and $(0\ 0.66\ 5)$ reflections below \hcb\ and 
$(0\ 0.25\ 4)$ and
$(0\ 0.5\ 4)$ above \hcb\ as function of magnetic field \hpb\
measured at a sample temperature of $T=4$~K. }
\end{figure}

The application of magnetic field at low temperature for the \hpb\
configuration results in a significantly different behavior of \ttb\ and
\tmn\ compared to the \hpa\ configuration. Here the sample was cooled
to 2~K in zero field and then data were measured with increasing field
up to $\mu_0\hpa=10$~T, (Fig. \ref{fig:hdep_2_10_20}a). Between an
applied field of 1 and 2~T, a phase transition is observed in the Tb
sublattice with a shift of the peak positions from the incommensurate
value $\ttb=0.43$\bstar\ to the commensurate value $\ttb=1/3$\bstar\ and
accordingly from 2$\ttb=0.86$\bstar\ to 2$\ttb=2/3$\bstar. 
Although this transition is only related to
the Tb order, the intensities of the Mn reflections are influenced by
this transition, whereas the wave vectors are not affected. As shown
in Fig.~\ref{fig:hdep_2k_int}, the intensities of the $(0,\ 0.55,\ 3)$
(A-mode) and the $(0,\ 0.55,\ 4)$ (F-mode) reflections show a distinct
behavior. While the $(0,\ 0.55,\ 3)$ intensity decreases at \hctb, the
$(0,\ 0.55,\ 4)$ intensity increases.

\begin{figure}
 \includegraphics[width=8cm]{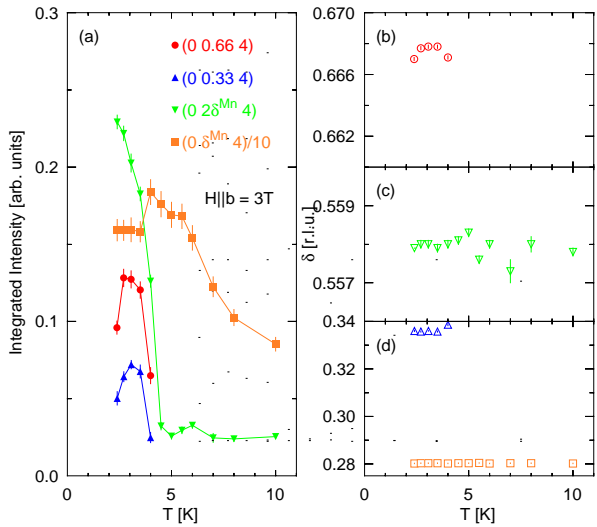}
 \caption{\label{tdep_3t} (Color online) (a) Temperature dependence
of the intensities of the Mn and Tb first and second harmonic 
superlattice reflections measured at magnetic field 
$\mu_0 H=3$~T. 
(b) shows the corresponding k-components of the wave vector.
The measurement is performed by going from the
intermediate Tb ordered phase through \tntb\ into 
the purely Mn ordered phase.}
\end{figure}

Cooling the sample in an applied field $\mu_0H=3$~T we find in the
temperature dependence that the transition to the incommensurate phase
with $\ttb=0.43$\bstar\ is suppressed at $\Tntb=4.5$~K with decreasing
temperature in preference for the commensurate $\ttb=1/3$\bstar\
phase (Fig.~\ref{tdep_3t}). For the F-mode reflections (0, \qmn, 4)
and (0, 2\qmn, 4) we find no changes in the value of \qmn\ but again a
clear change in the intensity is observed at \tntb.

\begin{figure}
 \includegraphics[width=8cm]{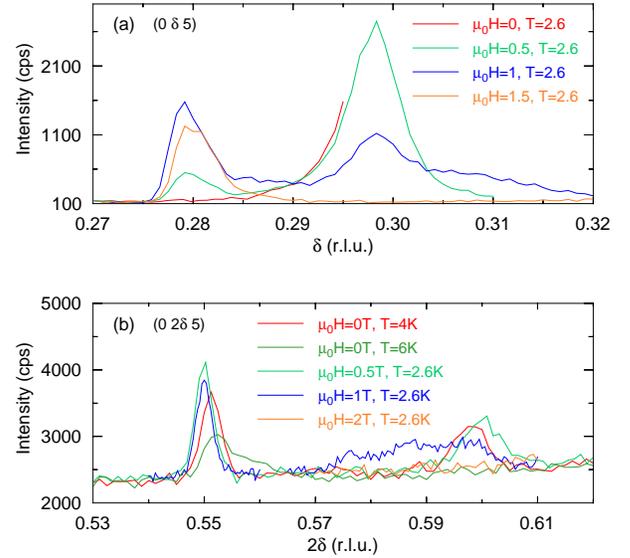}
 \caption{\label{peaks0306}
(Color online) 
k-scans over (a) (0, $\delta$, 5) and (b) (0, $2\delta$, 5) 
positions as function of
magnetic field \hpb\ and temperature below \tntb.}
\end{figure}

In this field-cooled measurement below $T=6$~K we find two previously
unreported modulations at (0, 0.298, 5), (0, 0.6, 5).  The nature of
these reflection is not clear at this point, however their field
dependence indicates that they are associated with Tb spin
ordering. We find that increasing field and increasing temperature
result in a considerable decrease in their intensity
(Fig.~\ref{peaks0306}a-b). The reflections disappear at the transition
to the \qtb=1/3 phase. Since these reflections only occur at the
high-energy x-ray setup they can be regarded as a bulk property
related to Tb magnetic order below \tntb.

\subsection{High-Field Commensurate Phase}

We now turn our attention to the commensurate phase above \hcb.  The scattering
geometry that allows us to measure the commensurate phase makes
only F- and C-modes accessible (Fig.~\ref{fig:hdep_10k_new}).  At $T=10$~K the
transition to a commensurate phase occurs at $\mu_0\hcb=9$~T.  Above \hcb, the
F- and C- modes exhibit a linear increase of their intensity with
field for the first harmonic reflections, whereas the intensity of the
second harmonic peaks remains constant with increasing field. This
behavior follows on from the lower field behavior in the incommensurate phase. As
the F- and C- modes arise from Tb-ordering it only allows us to
comment on the value of \tmn\ and not to changes associated with the
flop of the Mn spin-spiral. Cooling the sample to $T=2$~K and applying field 
we find the transition to the
commensurate phase at $\mu_0\hcb =5$~T. Here reflections associated with the $\ttb=1/3$\bstar\
phase vanish at \hcb\ and only commensurate reflections with $\qmn=1/4$ and
$2\tmn=1/2$\bstar\ are observed (Fig.~\ref{fig:hdep_2_10_20}). The integrated intensities of (0, 0.25, 4), (0, 0.5, 4) and (0, 0.75,
4) are plotted as function of \hpb\ in
Fig.~\ref{fig:hdep_2k_int}. Intensities of the 2\tmn\ reflection
are a factor of 30 weaker than the first harmonics.
While the intensities of the (0, 0.25, 4) and (0, 0.75, 4) reflections
increase linearly with field, the intensity of the (0, 0.5, 4) remains
constant up to maximum field. The Tb (0, 0.66, 5) reflection also
shows constant intensity from its onset at $\mu_0H=2$~T up to \hcb.
At \hcb\ a little intensity is still present and coexists with the
reflections of the commensurate phase.

\begin{figure}
 \includegraphics[width=8cm]{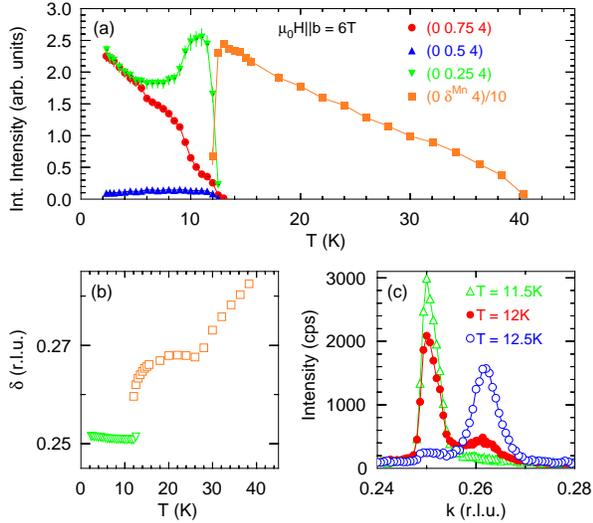}
 \caption{\label{tdep_6t} (Color online)
(a) Intensity behavior of CM and IC superlattice reflections 
measured at $H=6$~T
below (CM) and above (IC) the critical temperature. (b) Corresponding 
wave number $\deltamn$ as function of temperature for the first harmonic F-mode
reflection. (c) k-scans in the region of coexistence of the CM and IC
phase at the critical temperature.}
\end{figure}

Following the F- and C-mode reflections while cooling the sample in 
a field $\mu_0H=6$~T  we
find that \tmn\ first takes an incommensurate value which decreases with
lowering the temperature until $T=$\Ts\ (see Fig.~\ref{tdep_6t}a,b).
Below \Ts\ the value of \tmn\ remains relatively constant until
approximately 18~K where it rapidly decreases and locks into a value
of $\tmn=1/4$\bstar\ below 12~K. In Fig.~\ref{tdep_6t}c, k-scans in
the region of coexistence of the commensurate and incommensurate phase
at the critical temperature are shown. 
The variation in the intensity of the commensurate
reflections is quite different between the different reflections
we examined. Whereas the weak second harmonic reflection is constant in
intensity there is a strong variation for the first harmonic
reflections, indicating changes in the Tb order are obviously still
present even in the high field commensurate phase. 

\section{\label{discussion}Phase diagram and Discussion}

\begin{figure}
 \includegraphics[width=8cm,clip]{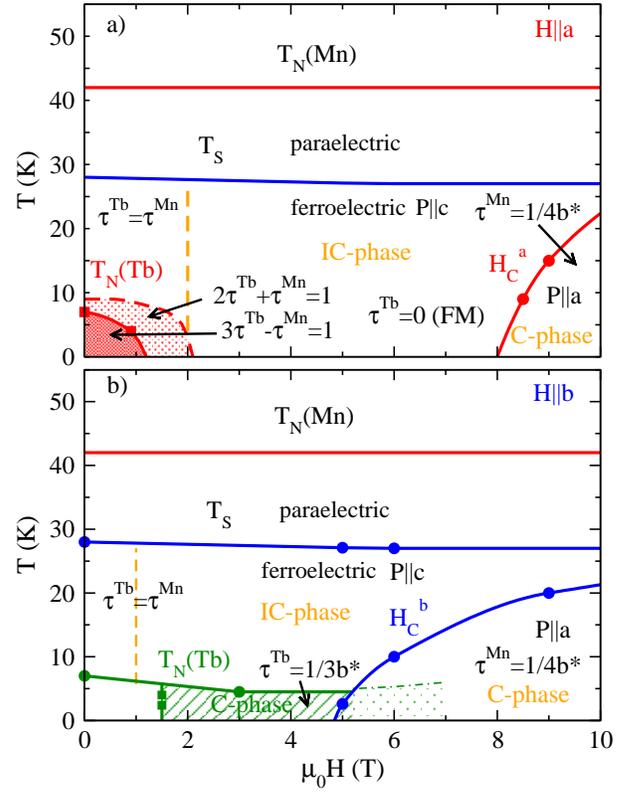}
 \caption{\label{phasediagram} 
(Color online) Phase diagram of TbMnO$_3$ for fields (a) \hpa\ and (b) \hpb. 
In (a) the light shaded area represents the intermediate phase region
and the dark shaded area the Tb AF ordered state found for field \hpa. 
In (b) the green line represents \tntb\ as
found for magnetic field \hpb. The green shaded area shows the intermediate 
CM Tb ordered phase for \hpb. The IC to CM transition for the
manganese order is shown for \hpa\ and \hpb.
The vertical dashed line defines the onset of the first harmonic reflections.
}
\end{figure}

From the measurements we have described in this paper we can construct
the $H-T$ phase diagrams for \hpa\ and \hpb\ (Fig.~\ref{phasediagram}a
and b) which summarize the various phases we have found and their
correlation to the ferroelectric polarization of TbMnO$_{3}$ reported
in Ref.~\onlinecite{Kim05}.

In zero field we find purely structural 2\tmn\ reflections which are
interpreted to arise from quadratic magneto-elastic coupling.  In the
SDW regime \tn$>T>$\Ts\ the intensity of the A-mode 2\tmn\ reflections
rises with cooling as expected for quadratic magneto-elastic coupling
and below \Ts\ it decreases as the Mn magnetic order becomes spiral.
While ideally 2\tmn\ reflections should disappear for a circular
spiral,\cite{Jia07} their continual observation for $T>\Ts$ reflects
the ellipticity of the Mn spiral ordering.

In the regime $\Ts>T>\ttb$ Mn- and and Tb-spin ordering is coupled
and \tmn=\ttb.\cite{Pro07b} Below \tntb, Tb-spins order with a
different wave vector but remain harmonically coupled to the Mn spiral
ordering as their wave vectors assume values that obey the relation
$3\ttb-\tmn=1$.  This is confirmed in this work to an accuracy of
0.003 and we depict the phase regions of 
this coupled ordering in Fig.~\ref{phasediagram}a
and b.

Application of magnetic field in this harmonically coupled state
results in different behavior depending on the direction of the field.
When a relatively small magnetic field is applied along the $a-$axis
below \tntb\ we find an intermediate state that is characterized by
the appearance of a Tb reflection at 0.9 (Figs.~\ref{tdep_0t_tb2}b and
\ref{hdep_4k_a}a), a jump of the wave vector \tmn\ to smaller values
and a shift of \ttb\ to 0.36\bstar\ observed by resonant magnetic
scattering (Fig.~\ref{hdep_4k_resonant}). The observed shifts in wave
vector are completely in agreement with the relationship $2\ttb+\tmn=1$
(Sections \ref{zerofield} and \ref{tbordered_hpa}) found in 
Ref.~\onlinecite{Pro07b}.  This region is marked as a light shaded
area in Fig.~\ref{phasediagram}a and appears both in the temperature
dependence at zero field (Fig.~\ref{tdep_0t_tb2}) between $\tntb=7$~K
and $\approx 9$~K as well as in the field dependence at $T=4$~K
between 1 and 2~T (Fig.~\ref{hdep_4k_a}). For $\mu_0\hpa>2$~T Tb spins
become ferromagnetically polarized. This intermediate phase may be visible 
only in the surface near region probed in this field configuration
and is likely to be absent if the true bulk of the crystal is probed.

It is worth noting at this point that the intermediate value of \ttb\
below \tntb\ for both \hpa\ and \hpb\ configurations is in fact quite
similar; \ttb=0.36\bstar\ in the former case and \ttb=0.33\bstar\ for
the latter. The difference in these two orderings of Tb-spins is that
for \hpa\ \tmn\ is harmonically coupled to the Tb-ordering while for
\hpb\ it is not.  This subtle but significant difference we believe
reflects the anisotropy of the Tb-spin ordering as well as the
delicate balance of the coupling between Mn and Tb spins.

When field is applied along the $b$-axis below \tntb\ the harmonically
coupled regime disappears and instead we find a phase transition of
the Tb moments at about $\mu_0H=1.25$~T from the incommensurate
to a commensurate order with $\ttb=1/3$\bstar\ and the 2nd harmonic
$2\ttb=2/3$\bstar\ (see Fig.~\ref{phasediagram}b).  This intermediate
Tb phase does not affect the wave vector \tmn, which only becomes
commensurate at $\mu_0\hcb\sim5$~T. Nevertheless, intensities are affected
which emphasizes again the effect of the Tb magnetic order on the
structure.  Similar to \hpa, a dashed vertical line in
Fig.~\ref{phasediagram}b shows the field of onset for the first order
satellites which are induced by the polarization of the Tb magnetic
moment by the magnetic field. As discussed before
(Section~\ref{tbordered_hpa}) in the Tb-ordered region $T<\tntb$ and
the intermediate region ($\mu_0H>0$), a clear assignment of Mn or Tb
reflections is no longer possible.

At higher field values for $\mu_0\hpa>2$~T and below \Ts\ only
incommensurate Mn reflections with wave vectors \tmn\ and 2\tmn\ are
observed, whereas the Tb reflections disappear.  Since first harmonic
reflections in field are found for G- and C-modes only and not for the
Mn-order related A-type reflections, this suggests that the vertical
dashed line in Fig.~\ref{phasediagram}a marks the onset of the FM
order of the Tb moments for applied fields \hpa. This also
corroborates neutron diffraction measurements which find the sharp
decrease of F- and C-modes in this field range and the sharp increase
of ferromagnetic Bragg reflections.\cite{Ali06} This FM ordering of
Tb-spins is relatively stable up to fields $\hpa>8$~T.

For both field configurations \hpa\ and \hpb\ we find that the flop of
the polarization is associated with a first order transition to an
commensurate $\tmn=1/4$\bstar\ phase. As the direction of the
polarization changes, this flop is taken to reflect the flop of the
spiral plane from the $bc-$ to the $ab-$plane.  The jump to a
commensurate wave vector at the spiral flop transition is not
surprising given that \qmn\ is in the proximity of a commensurate
value in zero field and therefore it is likely an energetically
favorable state can be reached by a relative small shift to
$\qmn=1/4$. This is in contrast to DyMnO$_{3}$ were the $\delta^{Mn}$ is
not close to a commensurate value ($\tmn=0.38$\bstar) and at the
polarization flop transition the wave vector does not shift at
all.\cite{Str07} In \tbmno\ above \hcb, variations in Tb order seem
still to be present as can be deduced from the strong variation of the
first harmonic Mn reflections as function of temperature at a field of
6~T shown in Fig.~\ref{tdep_6t}. Here the transition of Tb ordering
into the $\ttb=1/3$\bstar\ phase seems to extend into the commensurate
phase above \hcb, as is shown by the horizontal point-dashed line in
Fig.~\ref{phasediagram}b.

In more general terms the flop of the polarization from \ppc\ to \ppa\
can be understood by the magnetic degrees of freedom of the Mn
transverse spiral.  There are three excitations of the spiral. One is
the phase of the spiral polarized within the
$bc$-plane. Two other components are polarized 
along the $a-$axis in zero field.\cite{Senff07} 
One of these components results in a
flop of the $bc$-spiral by a rotation around the $b-$axis while the
other results in a twist of the spiral by a rotation around the
$c-$axis.  All measurements thus far are consistent with the assumption
that field applied
either along the $a-$ or the $b-$axis couples to the former mode and
produces the flop of the spiral from the $bc-$ to the $ab-$ plane,
while field applied along the $c-$axis melts the spiral ordering.\cite{Argyriou07}

As this picture can describe the main flop of the polarization as well
as the magnetic excitations, this work has shown that the magnetic
interactions between Mn and Tb spins remain significant and may be
considered as a perturbation upon this model of the flop of the
polarization and the spiral.  The strength of he $J_{Mn-R}$
interaction can be gauged by noting that the flopping fields for the
$R$=Dy are lower than in the present case for TbMnO$_{3}$. This can be
understood as a weaker $J_{Mn-R}$ for the case of $R$=Dy. Indeed in
DyMnO$_{3}$ at low temperature Dy-spins become completely uncoupled
from Mn-spins and Dy orders with wave vector $\tau^{Dy}=1/2$\bs, a
behavior significantly different from the harmonic coupling we find in
TbMnO$_{3}$ below \Tntb.\cite{Pro07b}


\section{\label{conclusions}Conclusions}

In this paper we present an extensive study of the behaviour of
structural and magnetic superlattice reflections of \tbmno\ as
function of temperature and applied magnetic field.  Details in the
$H-T$ phase diagram are revealed by small changes in the magnitude of
wave vector and intensities. These findings help us explain the
changes in spontaneous polarization as function of temperature and field. 
The subtle difference in the Tb order wave vector observed 
at low temperatures for non-zero 
fields \hpa\ and \hpb\ reflects the anisotropy of 
the Tb-spin order and the delicate balance between Mn and Tb spins.
This again highlights the significance of interactions between Mn and Tb spins
for the description of the polarization flop at the critical fields.
Differences in the magnetic wave vector in the crystal bulk and the
surface near region, as is observed by high-energy (100~keV) and hard x-ray 
investigations, show the influence of surface effects on magnetic 
polarization behaviour in multiferroic compounds.

\begin{acknowledgments}
We would like to thank W. Caliebe and C.S. Nelson for their assistance
at the experiment at NSLS. Work at Brookhaven was supported by the
U.S. Department of Energy, Division of Materials Science, under
Contract No. DE-AC02- 98CH10886. SL and DNA were supported by the 
Deutsche Forschungsgemeinschaft under contract AR-613/1-1.
\end{acknowledgments}


\end{document}